\documentclass[manuscript]{aastex}

\def \ms{m\,s$^{-1}$\,}
\def \kms{km\,s$^{-1}$}
\def \msun{$M_{\odot}$}
\def \rsun{$R_{\odot}$}
\def \lsun{$L_{\odot}$}

\def \mearth{$M_{\oplus}~$}
\def \me{$M_{\oplus}$}

\def \mj{$M_{\rm Jup}$}

\def \msini{$M\sin{i} ~$}
\def \rhk{$\log{R'_{\rm HK}}$}
\def \starname{HD~1461}                                 
\def \altnames{HIP~1499, HR~72, GJ~16.1, GJ~9009, BD~-0838, SAO~128690, SPOCS~13}
\def \spectype{G0V}                                     
\def \vmag{6.46}                                        
\def \starmasssou{$1.018\,\pm\,0.1$}       
\def \starmasstak{$1.026_{-0.030}^{+0.040}$} 
\def \starmassvaf{$1.08\,\pm\,0.04$}         
\def \starmass{1.022}                      
\def \starradtak{$1.11\,\pm\,0.04$}          
\def \starradvaf{$1.095\,\pm\,0.026$}        
\def \starlumsou{1.188\,$\pm$\,0.017}          
\def \starlumvaf{1.197\,$\pm$\,0.113}            
\def \teffsou{5765\,$\pm$\,18}                        
\def \teffcen{5808}                                   
\def \loggsou{4.38\,$\pm$\,0.03}                      
\def \loggcen{4.39}                                     
\def \loggtak{4.37\,$\pm$\,0.03}                        
\def \rhkhalla{$-5.04$}                                 
\def \rhkhallb{$-5.00$}                                 
\def \rhkjtw{$-5.03$}                                   
\def \rhkours{$-5.00$}                                  
\def \expjit{1.59 \ms}                                  
\def \agejtw{6.3}                                   
\def \agevaf{$4.2_{-2.1}^{+1.7}$}                   
\def \agetak{$7.12_{-1.56}^{+1.40}$}                
\def \vsini{1.6}                           
\def \prot{29}                                     
\def \fehsou{0.19\,$\pm$\,0.01}                         
\def \fehcen{0.20\,$\pm$\,0.01}                         
\def \fehvaf{0.18}                                      
\def \distance{23.4\,$\pm$\,0.5}                     
\def \DTRVy{12.8 years}                                 
\def \DTRVd{4687 days}                                  
\def \ndataRV{164}                                      
\def \medunc{0.72 \ms}                                  
\def \rmsmean{3.80 \ms}                                 
\def \DTPHOTy{12.2}                               
\def \ndataPHOT{799}                                    
\def \telescope{Keck}
\def \spectrograph{HIRES}
\def \fitepoch{JD 2450366.936}
\def \nplanets{two} 
\def \periodb{5.77} 
\def \periodd{4226} 
\def \periodc{446.1} 
\def \periodcfitthree{446.1} 
\def \perioddfitthree{5017} 
\def \periode{18.3} 
\def \FAPb{$1.1\times10^{-10}$}
\def \FAPd{$9.9\times10^{-16}$}
\def \FAPc{$4.5\times10^{-6}$}
\def \FAPe{$4.7\times10^{-5}$}
\def \minmassdjtwo{0.22}  
\def \minmassbeone{7.4} 
\def \minmasscethree{27.9}  
\def \minmassdethree{87.1}  
\def \ampbone{2.60 \ms} 
\def \ampdtwo{2.77 \ms} 
\def \ampcthree{2.30 \ms} 
\def \ampefour{1.49 \ms} 
\def \rmsb{3.43 \ms} 
\def \rmsd{2.87 \ms} 
\def \rmsc{2.41 \ms} 
\def \rmse{2.28 \ms} 
\def \transdepth{0.5} 
\def \transprob{8} 

\slugcomment{draft: hd1461\_rev3.6}

\shorttitle{A Super-Earth Orbiting \starname}
\shortauthors{Rivera et al.}

\begin{document}

\title{A Super-Earth Orbiting the Nearby Sun-like Star \starname}

\author{
Eugenio J. Rivera\altaffilmark{1},
R. Paul Butler\altaffilmark{2},
Steven S. Vogt\altaffilmark{1},
Gregory Laughlin\altaffilmark{1},
Gregory W. Henry\altaffilmark{3},
Stefano Meschiari\altaffilmark{1}
}

\altaffiltext{1}{UCO/Lick Observatory, University of California, Santa Cruz, CA 95064, USA}
\altaffiltext{2}{Department of Terrestrial Magnetism, Carnegie Institution of Washington, 5241 Broad Branch Road, NW, Washington, DC 20015-1305, USA}
\altaffiltext{3}{Center of Excellence in Information Systems, Tennessee State University, Nashville, TN 37209, USA}

\begin{abstract}

We present precision radial velocity (RV) data that reveal a Super-Earth mass planet
and two probable additional planets orbiting the bright nearby \spectype\ star
\starname.  Our \DTRVy\ of \telescope\ High Resolution Echelle Spectrometer precision RVs
indicate the presence of a \minmassbeone\,\mearth planet on a
\periodb-day orbit. The data also suggest, but cannot yet confirm, the presence
of outer planets on low-eccentricity orbits with periods of
\periodcfitthree\ and \perioddfitthree\ days, and projected masses (\msini) of
\minmasscethree\ and \minmassdethree\,\me, respectively.  Test integrations of
systems consistent with the RV data suggest that the configuration
is dynamically stable.  We present a \DTPHOTy-year time series of photometric
observations of \starname, which comprise \ndataPHOT\ individual measurements,
and indicate that it has excellent long-term photometric stability.  However,
there are small amplitude variations with periods comparable to those of the
suspected second and third signals in the RVs near 5000 and 446 days,
thus casting some suspicion on those periodicities as Keplerian signals.
If the \periodb-day companion has a Neptune-like composition, then its
expected transit depth is of order $d \sim$\transdepth\,mmag. The geometric
a priori probability of transits is $\sim$\transprob\%.
Phase folding of the ground-based photometry shows no indication that transits
of the 5.77-day companion are occurring, but high-precision follow-up of
\starname\ during upcoming transit phase windows will be required to
definitively rule out or confirm transits.
This new system joins a growing list of solar-type stars in the immediate
galactic neighborhood that are accompanied by at least one Neptune (or lower)
mass planets having orbital periods of 50 days or less.

\end{abstract}


\section{Introduction}

Over 400 extrasolar planets are now known.  The majority of these have been
discovered by using precision radial velocities (RVs) to detect the reflex barycentric
motion of the host star. We have had a large sample of over 1000 nearby stars
under precision RV survey for the past 13 years at \telescope\ with
the High Resolution Echelle Spectrometer (HIRES).  One of the target stars is \starname, a nearby
\spectype\ star, only \distance\ pc away \citep{per97}.  This star has been on
the \telescope\ program since 1996 October.  Over the past \DTRVy, we have
accumulated a total of \ndataRV\ precision RVs that indicate a
system of at least \nplanets\ planets orbiting this star. In this paper, we
present all of these RV data and discuss the planetary system that
they imply.

\section{Basic properties of the host star \starname}

\starname\ (\altnames) is a bright ($V$=\vmag) and well-studied star.
It has been characterized in a number of studies, including those of
\citet{vaf05}, \citet{takeda07}, and \citet{sou08}.  Table~\ref{stellarparams}
summarizes recent determinations of the fundamental stellar parameters for
\starname.  Taken together, these properties indicate that it is an old,
metal-rich, inactive star well suited for precision RV planet
searches.  Using Ca~H+K measurements taken between 1994 and 2006, \citet{hall07}
found that \starname\ is one of the 13 targets for which the observed variability is zero
within the uncertainties. In 51 observations over 8 seasons, they measured
a mean \rhk\ of \rhkhalla.  \citet{hall09} found a mean value of
\rhkhallb\ over seven seasons with seasonal averages ranging from $-4.96$ to
$-5.01$.  \citet{jtw04} found a mean value of \rhkjtw, along with an estimated slow
rotational period of \prot\ days and an age of \agejtw\ Gyr.  Our measurement of
\rhk=\rhkours\ leads to an estimate \citep{jtw05} of \expjit\ for the expected
RV jitter due to stellar surface activity. The age of
\starname\ was estimated as \agevaf\ Gyr by \citet{vaf05} and \agetak\ Gyr by
\citet{takeda07}.  The large discrepancy and uncertainties for these
chromospheric ages is not surprising since the correlation between age and
chromospheric activity becomes very weak for ages bigger than 2 Gyr (see e.g.
\citet{pp04}).  In summary, \starname\ is a nearby, bright star with physical
properties that are quite similar to our own Sun, and is an ideal candidate
star for the application of high-precision Doppler-velocity monitoring.

\section{Observations}

The \spectrograph\ \citep{vog94} of the \telescope\ I telescope was
used to monitor \starname.  A total of \ndataRV\ \telescope\ observations were
obtained, from 1996 October 10 to 2009 August 10, a data span of \DTRVd. The
median internal velocity uncertainty for these \telescope\ data is \medunc.

Doppler shifts were measured in the usual manner \citep{but96} by placing an
iodine absorption cell just ahead of the spectrometer slit in the converging
beam from the telescope.  This gaseous iodine absorption cell superimposes a
rich forest of iodine lines on the stellar spectrum, providing a wavelength
calibration and proxy for the point-spread function (PSF) of the spectrometer.
The iodine cell is sealed and temperature-controlled to 50 $\pm$ 0.1 ${^\circ}$C such that
the column density of iodine remains constant.  For the \telescope\ planet
search program, we operate the \spectrograph\ at a spectral
resolving power $R\approx70,000$ and a wavelength range of 3700-8000\,\AA,
though only the region 5000-6200\,\AA\ (with iodine lines) is used in the
present Doppler analysis.
The iodine region is divided into $\sim$700 chunks of 2\,\AA\ each.  Each chunk
produces an independent measure of the wavelength, PSF, and Doppler shift.  The
final measured velocity is the weighted mean of the velocities of the individual
chunks.

Table~\ref{vels} lists the complete set of \ndataRV\ RVs for
\starname, corrected for the solar system barycenter.  The table lists the barycentric JD of
observation center, RV, and internal uncertainty.  The internal
uncertainty reflects only one term in the overall error budget, and results
from a host of systematic errors from characterizing and determining the PSF,
detector imperfections, optical aberrations, effects of under-sampling the
iodine lines, etc.  Two additional major sources of error are photon statistics
and stellar jitter.  The latter varies widely from star to star, and can be
mitigated to some degree by selecting magnetically inactive older stars and by
time-averaging over the star's unresolved low-degree surface $p$-modes.  For most
of the past \DTRVy, only single exposures at \telescope\ were taken of
\starname\ at each epoch.  Since these single exposures were much shorter than
the characteristic time scale of low-degree surface $p$-modes on the star, they
suffered from additional noise (stellar jitter).  By 2008 July at \telescope,
we began $p$-mode averaging each observation, combining multiple shots of
\starname\ over a 5-10 minutes dwell at each epoch.  All observations have been
further binned on two-hour timescales.

Although all of the observations used in this work were obtained with the
\spectrograph\ at the
\telescope\ I telescope, 28 of the RVs in Table~\ref{vels} were
derived from publicly available spectra from the NASA Keck Observatory
Archive.  These velocities are marked with ``Q01'' in the observatory column
in Table~\ref{vels}.  The CCD format of the Q01 run was shifted by several
angstroms relative to our long-term standard, introducing a zero-point velocity
offset between Q01 and our standard set-up.  This is an additional parameter to
be determined in the fits discussed below.

\section{Photometry}

In addition to our RV observations. we acquired high-precision 
photometric observations of \starname\ during 13 consecutive observing 
seasons from 1996 November to 2009 January with the T8 0.80 m automatic 
photometric telescope (APT) at the Fairborn Observatory. Our APTs can detect 
short-term, low-amplitude brightness variability in solar-type stars due 
to rotational modulation of the visibility of surface magnetic activity 
(spots and plages), as well as longer-term variations associated with the 
growth and decay of individual active regions and the occurrence of stellar 
magnetic cycles \citep{henry99}. The photometric observations help us to 
establish whether observed RV variations are caused by 
stellar activity or planetary reflex motion \citep[e.g.,][]{hbd+2000}. 
\citet{queloz01} and \citet{paulson04} have presented several examples of 
periodic RV variations in solar-type stars caused by 
photospheric spots and plages. The photometric observations are also useful 
to search for transits of the planetary companions
\citep[e.g.,][]{hmbv2000,char2000}.

The T8 APT is equipped with a two-channel precision photometer that
separates the Str\"{o}mgren $b$ and $y$ passbands with a dichroic filter 
and takes simultaneous 30 s integrations with two electromagnetic interference (EMI) 9124QB bi-alkali 
photomultiplier tubes.  The APT measures the difference in brightness 
between a program star and a nearby constant comparison star or stars. 
The typical precision of a single observation is approximately 0.0015 mag,
as measured for pairs of constant stars.  The automatic telescopes, 
photometers, observing procedures, and data reduction techniques are 
described in \citet{henry99}.  Further details on the development and 
operation of the automated telescopes can be found in \citet{henry1995a,
henry1995b} and \citet{eaton03}.

For \starname, we used HD~2361 ($V=7.89$, $B-V=0.470$, F2) as our primary
comparison star. The individual Str\"{o}mgren $b$ and $y$ differential 
magnitudes have been corrected for differential extinction with nightly 
extinction coefficients and transformed to the Str\"{o}mgren system with 
yearly mean transformation coefficients.  Since \starname\ lies at a 
declination of $-8\arcdeg$, the photometric observations from Fairborn 
were made at airmass 1.3--1.8, which is somewhat higher than most
observations.  Therefore, to improve the precision of the measurements,
we combined the Str\"{o}mgren $b$ and $y$ differential magnitudes into 
a single $(b+y)/2$ passband.  

\begin{figure}
\epsscale{0.9}
\plotone{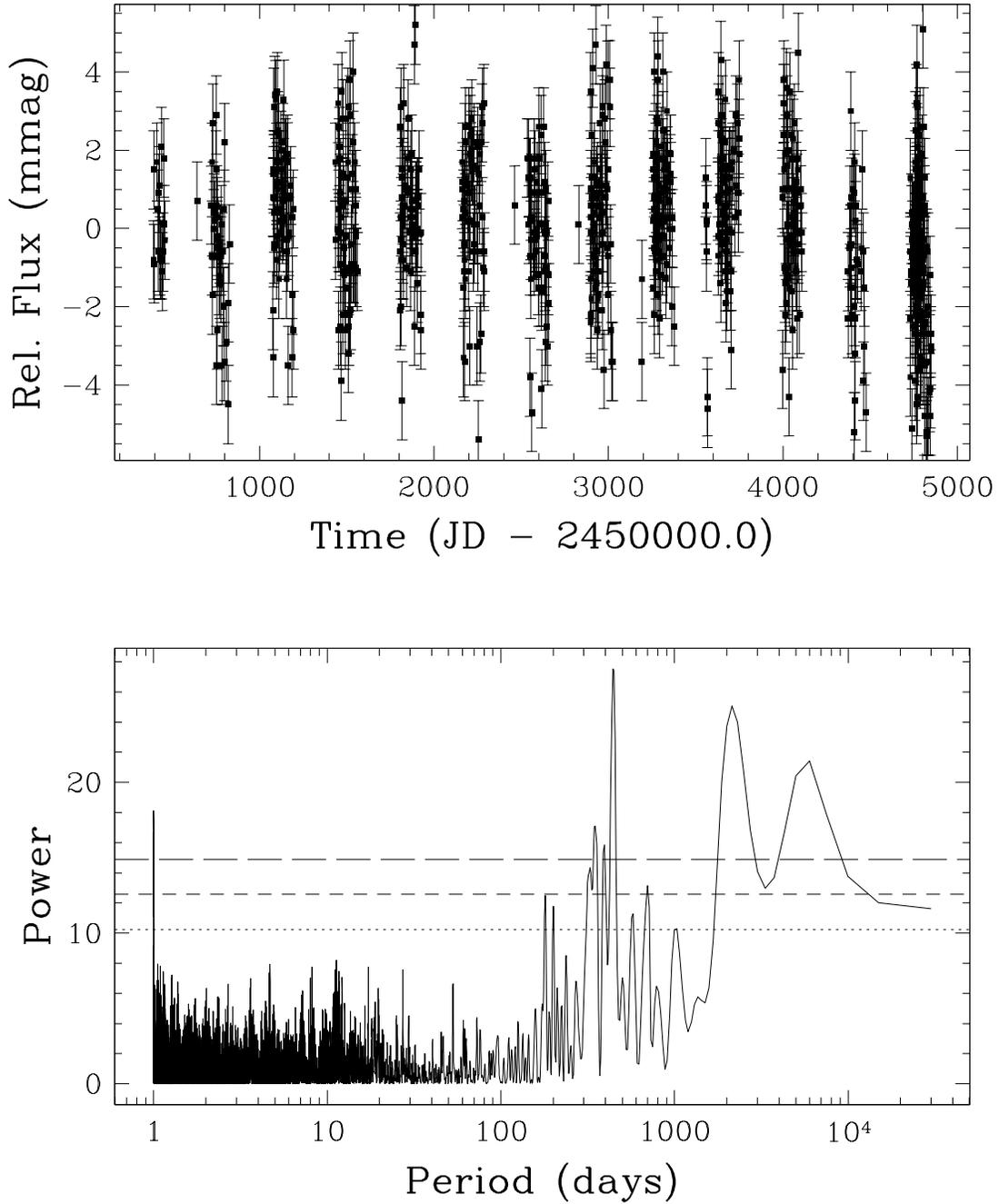}
\caption{Top: differential photometry of \starname.
Times are barycentric JD.
Bottom: Lomb-Scargle periodogram of the photometry.}
\label{photometry}
\end{figure}

A total of \ndataPHOT\ differential magnitudes from 13 observing
seasons are plotted in the top panel of Figure~\ref{photometry}.  The data 
scatter about their mean with a standard deviation of $\sigma=0.0019$~mag, 
which provides an upper limit to possible brightness variation in 
\starname.  The data are plotted such that observations that are
brighter/dimmer than the mean are indicated with positive/negative flux 
values.  A Lomb-Scargle periodogram of the photometric measurements is 
shown in the bottom panel of Figure~\ref{photometry} and reveals a 
significant periodicity within the data at 444.5 days with an estimated false
alarm probability (FAP) of $3.3\times10^{-9}$ (calculated using procedures
described in \cite{GB87} and \citet{cum04}).  This peak is near the tallest 
peak in the RV periodogram of the residuals of the circular 
two-planet fit discussed below.  A second peak spans from 3000 to 9000 days with 
an estimated FAP of $2.1\times10^{-6}$.  This broad peak is similar to 
the one present in the periodogram of the residuals of the (circular) 
one-planet fit also discussed below. The three horizontal lines in the bottom 
panel of Figure~\ref{photometry} and in all similar periodograms below 
represent FAPs of 0.1\%, 1\%, and 10\% from 
top to bottom, respectively.  We computed least-squares sine fits for the 
three RV planet candidates described below.  Sinusoid fits 
with periods \periodb, \periodcfitthree, and \perioddfitthree\ days yield 
semi-amplitudes of only 0.00018, 0.00070 and 0.00061 mag, respectively.
We conclude that the lack of significant coherent photometric variability
of \starname\ supports planetary reflex motion as the cause of the 5.77-day
periodicity in the RV measurements. The weak periodicities observed
at 445 and 3000-9000 days, on the other hand, do warrant some caution in
interpreting the RV periodicities at those periods as being due
to planetary companions.

\section{The Planetary System Orbiting \starname}

\begin{figure}
\includegraphics[angle=-90,scale=0.65]{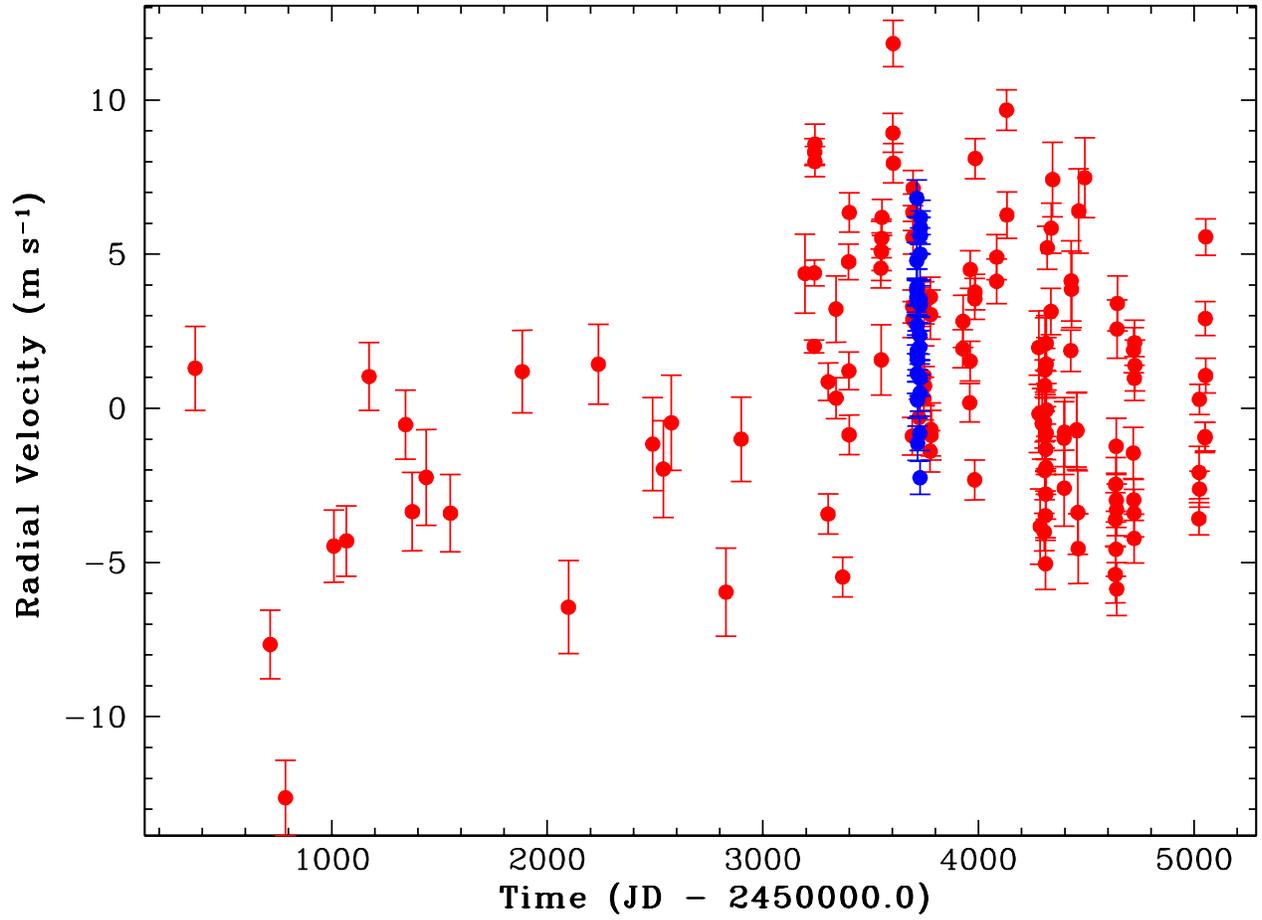}
\caption{Relative RVs of \starname.
Velocities based on spectra obtained by our group are in red,
and those based on archived Keck spectra (Q01) are in blue.}
\label{velocities}
\end{figure}

The RVs show a
rms scatter of \rmsmean about the mean velocity.  This
significantly exceeds the combined scatter due to the underlying precision
of both our measurement pipeline {\em and} the scatter expected in
this star due to its predicted \expjit level of stellar jitter.
Figure~\ref{velocities} shows the RV data set from Table~\ref{vels}.
An offset of -1.864 \ms (Q01 - Keck) has been applied between the two sets.
Note that in general, observations prior to 2004 August have larger internal
uncertainties.  The improvement in the internal uncertainties is a result of
the CCD upgrade discussed in some detail in \citet{riv05}.  Also, the
observations prior to 2004 August appear to be offset from those taken after
the CCD upgrade.  This is an artifact of our sampling with low frequency a
period that is not too distinct from one year during the first few years of our
observations.  As a result, during this time, we always sampled the period
of \starname\,c such that the star's RV is low.

\begin{figure}
\plotone{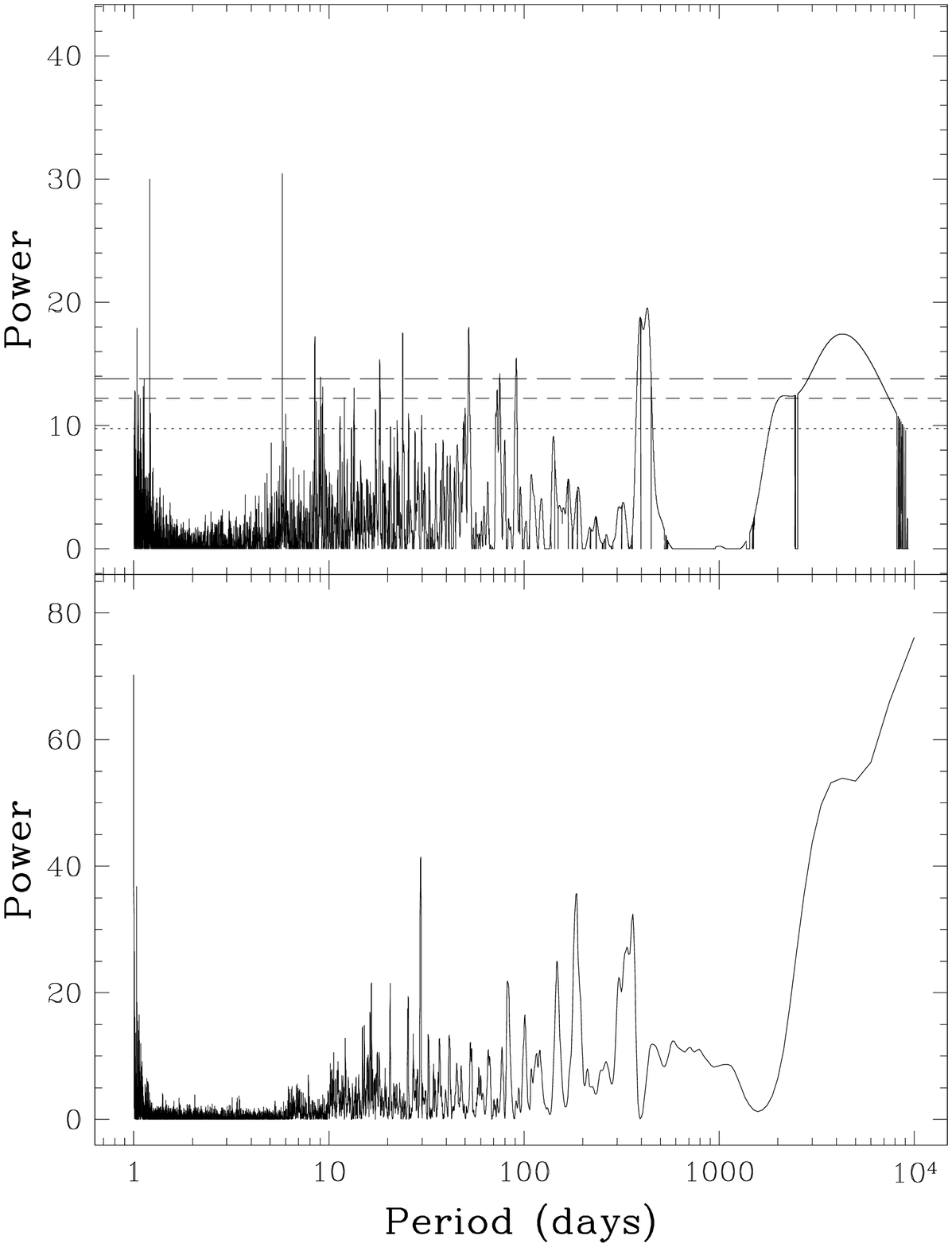}
\caption{Top: periodogram of the RV data
set for \starname.  The tallest peak is at 5.77 days.
Bottom: power spectral window of the RV data.}
\label{period0}
\end{figure}

Figure~\ref{period0} shows the periodogram of the RV
data set (top panel).  Power at each sampled period is proportional to
the relative improvement (drop in $\chi_{\nu}^2$) in the fit quality for
a circular model versus a constant velocity model. The
periodogram shows a number of significant signals, with the strongest peak
occurring at a period of \periodb\ days. The FAP of
this peak is estimated (adopting the procedure described in \cite{cum04}) to be
\FAPb. Furthermore, this \periodb-day signal lies well away
from the periods favored by the sampling window (Figure~\ref{period0}, bottom
panel), which produces spurious power at periods near 29.6, 186.3, 361.5,
147.8, and 82.4 days. Signals near any of these periods would be suspected of
being artifacts of the observing scheduling.

\begin{figure}
\plotone{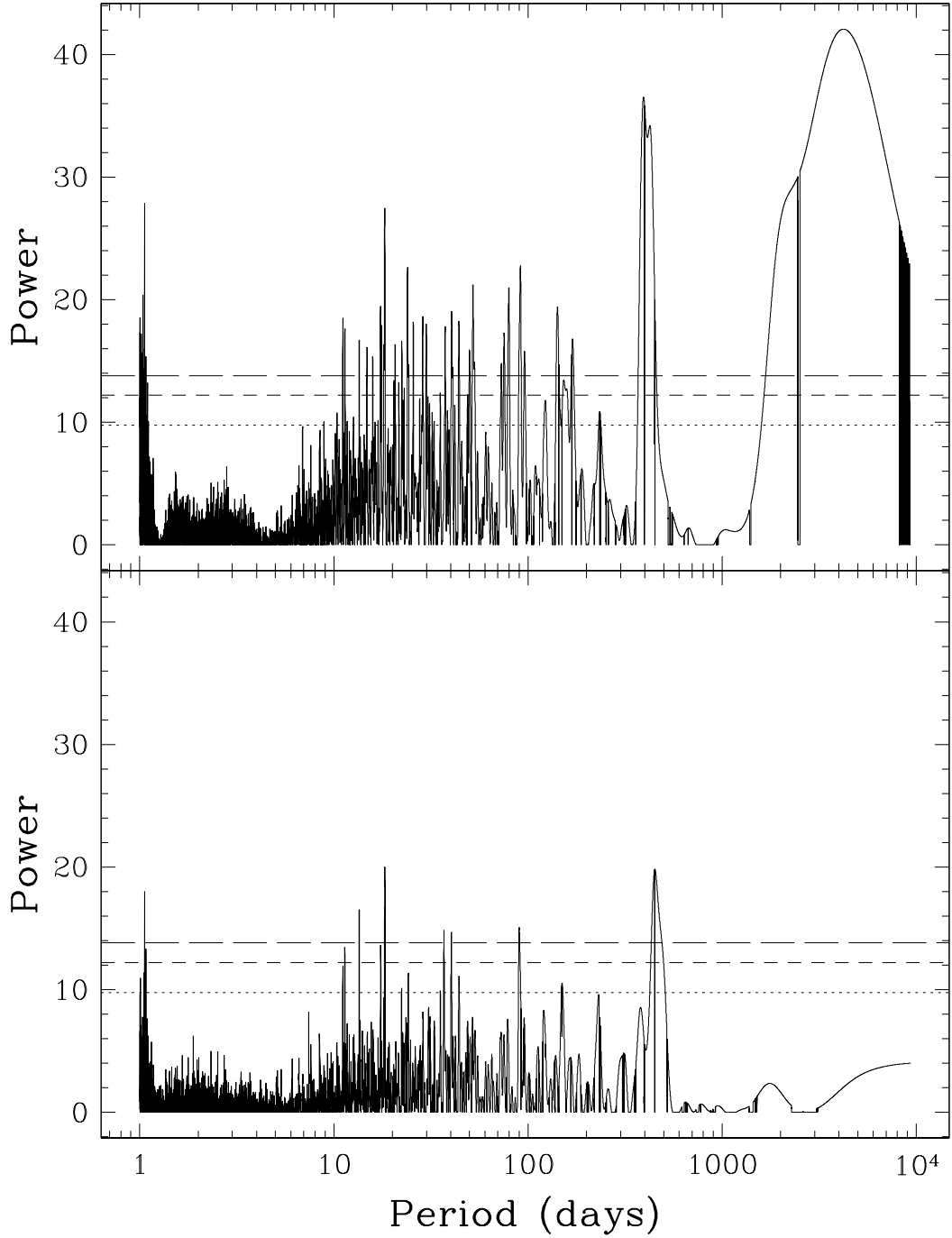}
\caption{Top: periodogram of the one-planet residuals,
assuming a circular orbit, of the RV data set for \starname.
Bottom: periodogram of the two-planet residuals,
assuming circular orbits.}
\label{period12}
\end{figure}

Based on the periodogram, with an assumed stellar mass of \starmass\,\msun, we
fit a planet of mass \minmassbeone\,\mearth and period \periodb\ days on a
circular orbit to the RV data. The presence of this planet (with a RV
semi-amplitude of $K$\,=\,\ampbone) reduces the rms scatter of
the velocity residuals to \rmsb. Figure~\ref{period12} (top panel) shows the
one-planet residuals periodogram which has a strong peak at \periodd\ days.
The \periodd-day signal has a FAP of \FAPd\ and can be modeled with a
companion with $K$\,=\,\ampdtwo\ and a mass of \minmassdjtwo\ \mj.
The addition of this planet further reduces the rms scatter to \rmsd.

Figure~\ref{period12} (bottom panel) shows the two-planet residuals periodogram,
assuming circular orbits, which has a peak at a period of \periodc\ days. The
FAP of the \periodc-day signal is \FAPc.  If Keplerian, this periodicity can be
ascribed to the presence of a \periodcfitthree-day,
\minmasscethree\,\mearth companion.  This third component has
$K$\,=\,\ampcthree. The three-planet model, assuming circular
orbits, has an rms scatter of \rmsc.

\begin{figure}
\includegraphics[angle=-90,scale=0.65]{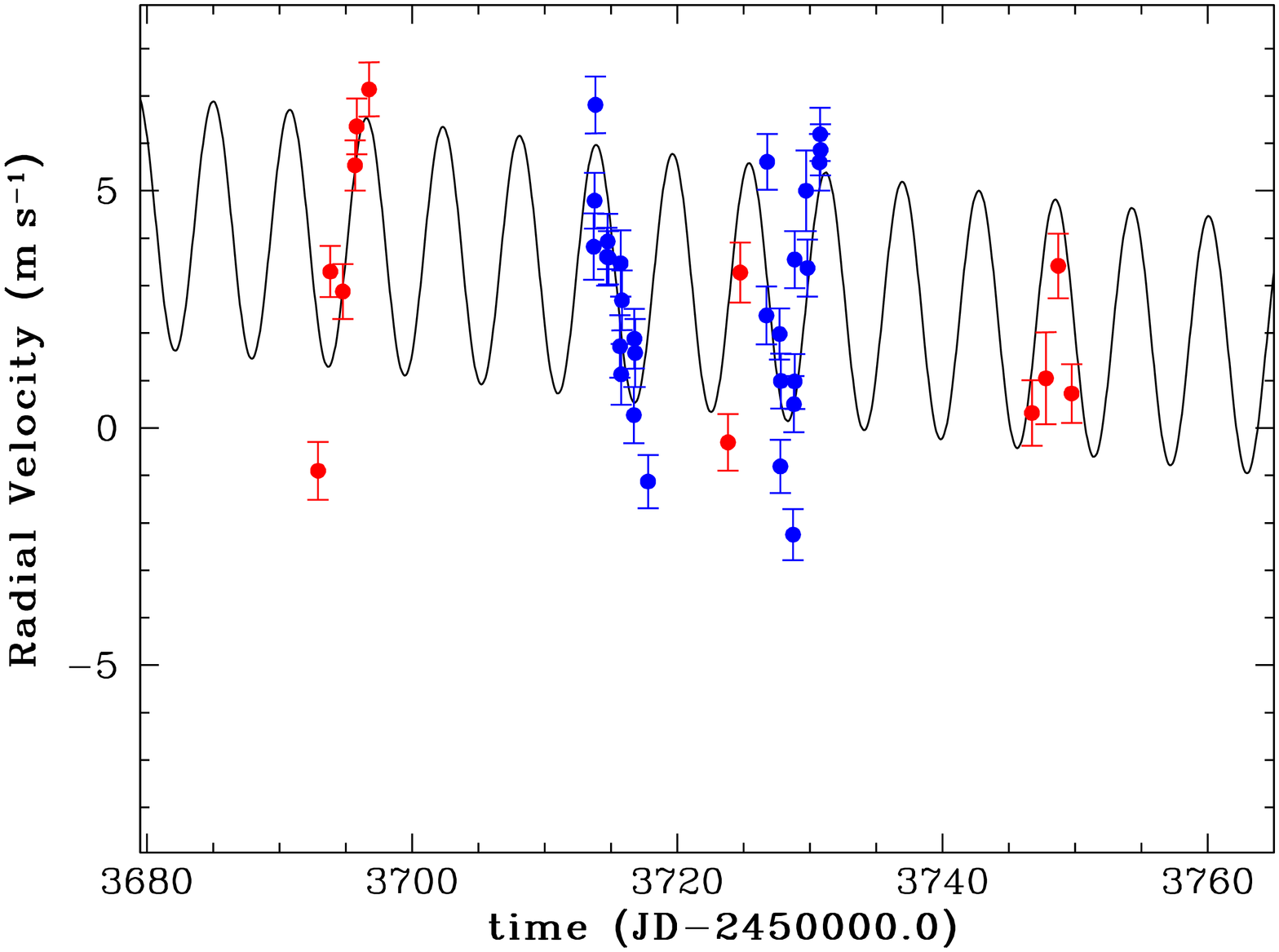}
\caption{Zoomed-in view of the circular three-planet model for \starname\,(black
curve).  Overplotted are the Keck observations in red and the Q01 observations
in blue.}
\label{rvmodzoomin}
\end{figure}

Figure~\ref{rvmodzoomin} shows the RV model for the three-planet
fit with circular orbits with the observations overplotted.  It is centered
around the time of the high cadence Q01 observations.

Given the one-, two-, and three-planet models, we can look either for solutions
in which the planetary orbits are circular or solutions where the eccentricities
are allowed to float. Inclusion of eccentricities provides only a modest
improvement to the orbital fits, leading us to conclude that a significant
amount of additional Doppler velocity monitoring will be required to improve the
eccentricity uncertainties.  However, allowing the eccentricities to float for
the two-planet fit dramatically reduces the significance of the 446-day
periodicity observed in the two-planet residuals (see below).
Additionally, if we fit for a second planet with period in the range
$\sim$390\,--\,450 days on an eccentric orbit, the significance of the
long-period planet can also be reduced.
In Tables~\ref{planetparamsnoe} and \ref{planetparamswithe}, we present our
best-fit versions of the system under the assumption of circular orbits
(Table~\ref{planetparamsnoe}) and with the additional degrees of freedom
provided by fully Keplerian trajectories (Table~\ref{planetparamswithe}). For
the orbital fits, we assume $i=90^{\circ}$ and $\Omega=0^{\circ}.$ The inclusion
of planet-planet gravitational interactions in the fits were found to be
unnecessary.  Uncertainties are based on 1000 bootstrap trials following the
procedure in Section 15.6 from \citet{pre92}. The standard deviations of the
fitted parameters to the bootstrapped RV's were adopted as the uncertainties.
The fitted mean anomalies are reported at epoch \fitepoch.  The mass of the host
star is assumed to be \starmass\,\msun, the mean of the isochrone masses of
\cite{sou08} (\starmasssou\,\msun\,) and
\cite{takeda07} (\starmasstak\,\msun\,).  Our fitting was carried out with the
publicly available {\it Systemic Console} \citep{systemic}.

\begin{figure}
\plotone{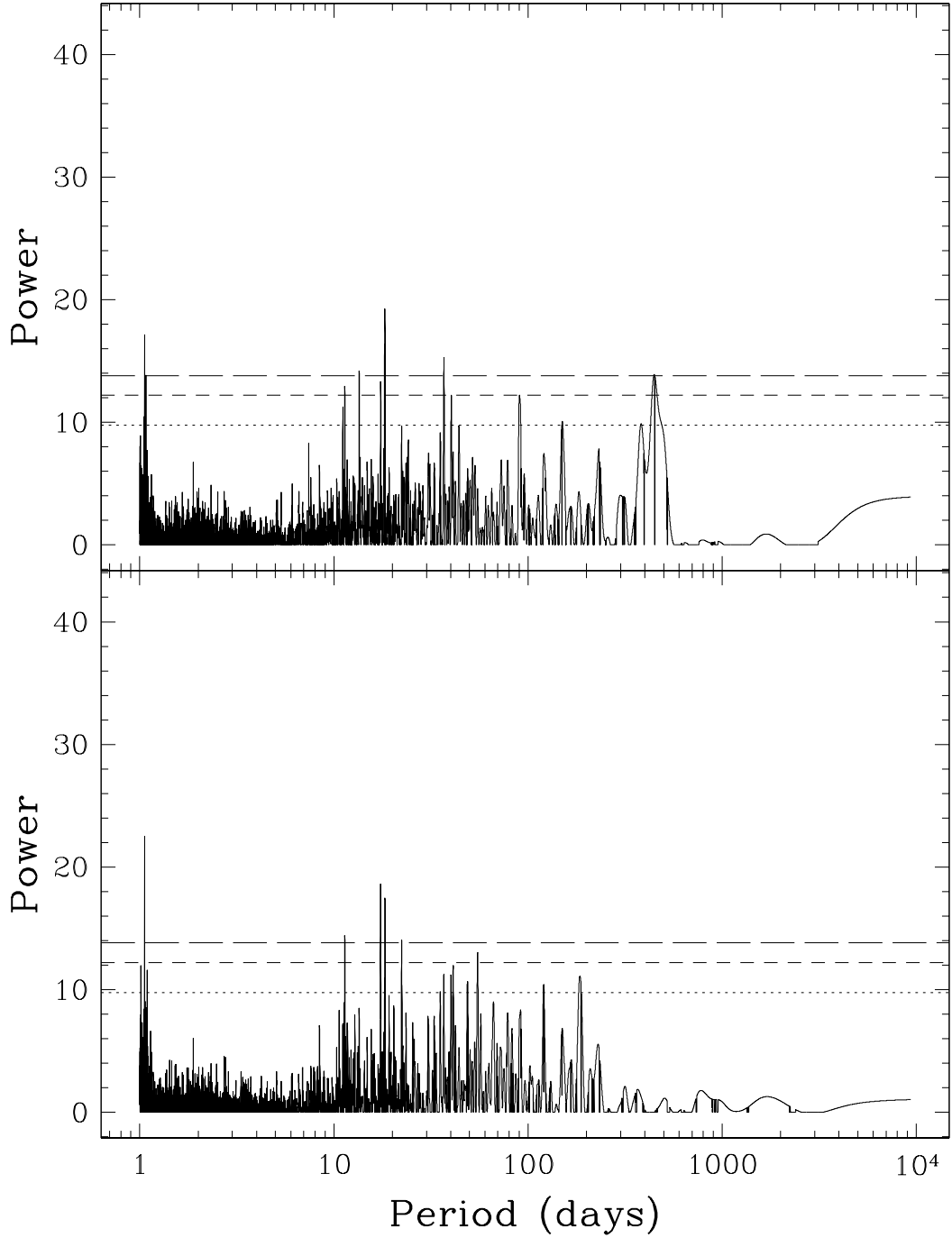}
\caption{Top: periodogram of the two-planet residuals,
with floating eccentricities, of the RV data set for \starname.
Bottom: periodogram of the three-planet residuals,
assuming circular orbits.}
\label{period23}
\end{figure}

Figure~\ref{period23} (top panel) shows the power spectrum of the velocity
residuals for the two-planet fit with floating eccentricities.
It is clear that the significance of the \periodc-day planet is dramatically
reduced.

Figure~\ref{period23} (bottom panel) shows the power spectrum of the velocity
residuals for the three-planet, circular fit.  There is a peak near
\periode\ days with a FAP of \FAPe.  Although using this period in a four-planet
circular fit results in a significant improvement in $\chi_{\nu}^2$, the rms
decrease from \rmsc\ for the three-planet model to \rmse\ is not significant.
Additionally, the fitted amplitude of \ampefour\ is significantly smaller than
the scatter around the model.

\begin{figure}
\epsscale{0.75}
\plotone{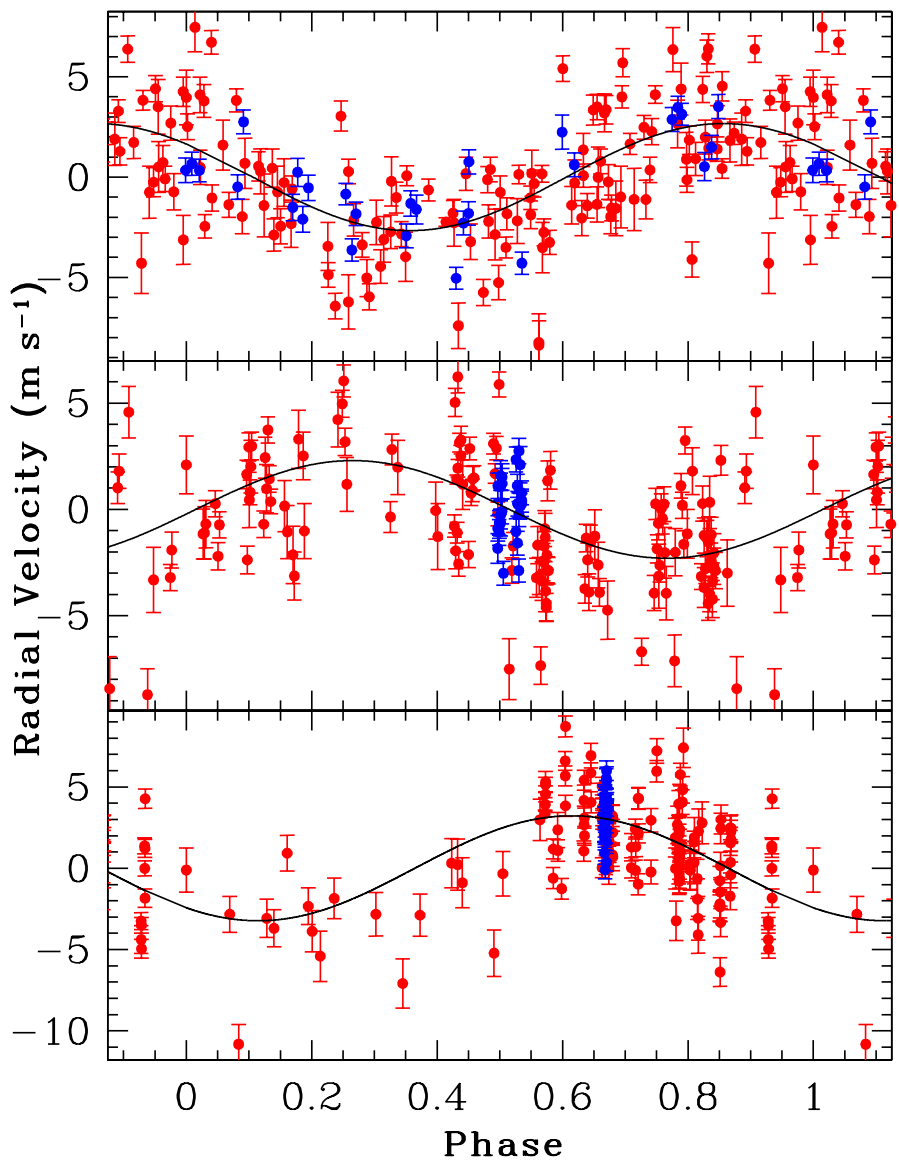}
\caption{Top: radial velocity of \starname\ due to planet b folded at
\periodb\ days.
Center: radial velocity of \starname\ due to planet c folded at
\periodcfitthree\ days.
Bottom: radial velocity of \starname\ due to planet d folded at
\perioddfitthree\ days.
In each panel, the effect of the other two planets has been subtracted out.
The curves represent the model velocities due to each respective planet.
The \telescope\ observations are shown in red, and the Q01 velocities are
shown in blue.
}
\label{phasedplot}
\end{figure}

In summary, the \telescope\ \spectrograph\ RV data show strong
evidence for at least two planets in orbit about the \spectype\ star \starname.
One of these has a short period of \periodb\ days. The other has a large period
of about 4000\,--\,6000 days, with a large uncertainty in both its period and
eccentricity. There are also hints of a third planet in the system near 446 days,
but the large uncertainties in the parameters of the second planet cast some
doubt on the presence of a third planet in the system.  Also, if we choose to
fit for an eccentric planet with a period of $\sim$390\,--\,450 days, we also find
large uncertainties in its parameters.  This would cast some doubt on the
presence of the large period planet.  The close match between
the photometric and the Doppler periodicities near 445 days leads us to tread
carefully in ascribing a planetary origin to our 446-day RV signal.
We note, however, that a 446-day period is too long to be associated with the
stellar rotation, given the star's measurable rotational velocity component,
$V\sin{i}$\,=\,\vsini\,\kms. If the 446-day Doppler signal does arise from
stellar activity, then the periodicity would need to be primarily associated
with the lifetimes of the active regions rather than with modulation induced by
stellar rotation.  Further observations will be needed to refine the parameters
of the second planet before a definitive detection claim can be made for a third
planet. That said, in Figure~\ref{phasedplot} we show the barycentric reflex
velocity of the host star due to each of three individual companions on circular
orbits in the system. In each panel, the velocities are folded at the period of
each corresponding planet.

\section{Dynamical Analysis}

It is useful to verify, via numerical integration, whether the planetary
configurations listed in Tables~\ref{planetparamsnoe} and
\ref{planetparamswithe} are dynamically stable. Such an analysis is particularly
useful in giving rough bounds on the allowed coplanar inclinations relative to
the line of sight to Earth. For simulations that do not include the effects of
tidal dissipation, Newtonian parameters were used for the initial states for
long-term integrations. The MERCURY integration package \citep{cha99} was used
for the simulations with a time step of 0.1 day. The first order post-Newtonian
term in the star's potential was also included, as in \cite{lis01}.

If the three orbits are assumed to be initially circular, with periods, masses
and mean anomalies given in Table~\ref{planetparamsnoe}, the system is stable
for at least 20 Myr.  Additionally, assuming the system to be coplanar, if we
set the inclination to the sky plane to various values from $i=90^{\circ}$ all
the way down to $i=1^{\circ}$ and perform a Newtonian fit for the other nine
parameters (three parameters per planet plus the two offsets), $\chi^2_{\nu}$ does
not change significantly from the nominal $i=90^{\circ}$ fit.  We also find the
$i=1^{\circ}$ fit to result in a system that is stable for at least 20 Myr.  For
this inclination, the fitted masses exceed 1.4, 5.1, and 16 $M_{\rm Jup}$.  This
system is stable because of the small eccentricities.  Thus, under the
assumption that the system is coplanar and the orbits are (nearly) circular, we
cannot place a lower bound on the inclination of the system.

The parameters of the floating-eccentricity version of the \starname\ system
given in Table~\ref{planetparamswithe} were also used as the initial input
conditions for a $10^7$ yr simulation.  The three-planet configuration is
disrupted in less than 1.4 Myr.

If additional RV measurements point to secure non-zero
eccentricities for the \starname\ planets, then studies of the long-term
dynamical evolution of the system should take the possibility of tidal
dissipation in planet b into account.

\section{Check for Transits by Companion b}

We performed a simple search for transits of planet b in the
\starname\ photometry.  The RV's determine the period of the inner planet to be
\periodb\ day.  The corresponding mean anomaly of the planet, likewise, is
$M= 79.045^{\circ}$.  The stellar radius is taken to be 1.1 $R_{\odot}$, the
mean of the values in Table~\ref{stellarparams}.

If \starname\,b's orbital plane allows for transits, then $\sin{i_b}\sim 1$, and
hence $M_{b}= 7.4\,M_{\oplus}$. If we assume that the planet migrated inward from
beyond the ice line, its composition is likely dominated by water. The models
of \citet{fortney07} suggest an $R= 2.7\,R_{\oplus}$ radius for such a planet,
leading to a central transit depth of $d\sim 0.05\%$, or $d\sim 0.5$ mmag.
If the planet has a massive atmosphere, its transit depth will be larger still.

At the time of the first photometric data point, $T_{p1}= {\rm JD}2450393.7339$,
our orbital model indicates a mean anomaly $M= 310.255^{\circ}$ for planet b.
For a circular orbit, transits occur at $M=90^{\circ}$. This means that relative
to phase $\phi=0.0$ referenced to $T_{p1}$, transits are centered at
$\phi=0.388$. The planet's orbital velocity is $v_{b}\sim 120$ \kms\ which, for
our $R= 1.1\,R_{\odot}$ star, implies a transit duration of
$\tau= 12000\,{\rm sec}\,= 3.4\,{\rm hours}$.  Assuming central transits, planet b
would be observable in transit 2.4\% of the time.

We assumed a simple ``top-hat'' step-function model for the transit, with the
depth, period, and phase given above. The photometric data set contains
\ndataPHOT\ individual measurements, with $\sigma=0.00188$~mag. Phase coverage
at the 5.77-day period of planet b is excellent. One thus expects $N=19 \pm 4$
points to lie in the transit window, indicating that, given the data set and the
presence of a transiting planet b, one can expect to detect a transit with
signal-to-noise ratio, S/N$\sim$1. We therefore conclude that the present photometric
data set is insufficient to make a definitive call as to whether transits are
occurring. Follow-up with high-precision, high-cadence photometry from either
ground or space is therefore warranted.

\section{Discussion}

We present evidence for at least two (and possibly three) low-mass planets
orbiting the nearby star \starname. With a RV semi-amplitude
$K=2.8\pm0.3$ \ms, the $M\sin{i}=7.8\pm0.8\,M_{\oplus}$ inner planet
\starname\,b is among the very lowest-amplitude companions yet detected using
the Doppler velocity technique.

The \starname\ system is thus another nearby case that joins the emerging
population of planets postulated by \citet{mayor40307}, who inferred that about
30\% of solar-type stars in the immediate galactic neighborhood are accompanied
by Neptune (or lower) mass planets having orbital periods of 50 days or less. 

With a period of only 5.77 days, \starname\,b has a non-negligible $P\sim8$\%
probability of transiting its parent star. While our phase-folded ground-based
photometry does not have the requisite cadence and S/N to detect such transits,
it would be readily possible to determine whether transits occur by making
high-cadence, high-precision observations that span the transit window
(see, e.g. Johnson et al.\ 2009). \starname\,b's mass is very similar to that of
CoRoT-7b, which has recently been determined to have a density similar to that
expected for a rocky planet (Queloz et al.\ 2009).  It would be of great
interest to learn whether \starname\,b is similarly dense, or whether its
composition is more reminiscent of ice-giant planets such as Neptune,
Gliese 436b, and HAT-P-11b.  Such a transit could be detected from space using,
for example, the Warm Spitzer platform.

As RV data bases grow in the monitoring of chromospherically quiet,
nearby stars, systems like \starname, are becoming increasingly common. With
each new system, the evidence is growing stronger that Super-Earths and other
low-mass planets are common around nearby Sun-like and cooler stars.  It is thus
only a matter of time and adequate cadence before Super-Earth planets are found
in the habitable zones of nearby stars. This is only a first reconnaissance of
this fascinating and quite nearby system. As RV data bases grow for
this star, the orbital ephemerides of these planets will become better
determined, and more planets will probably be revealed.

\acknowledgments

We acknowledge the major contributions of fellow members of our previous
California-Carnegie Exoplanet team: Geoff Marcy, Debra Fischer, Jason Wright,
Katie Peek, and Andrew Howard, in helping us to obtain the RVs
presented in this paper. We thank the anonymous referee who helped improve this
work.  S.S.V. gratefully acknowledges support from NSF grant
AST-0307493, and from the NASA KECK PI program.  R.P.B. gratefully acknowledges
support from NASA OSS Grant NNX07AR40G, the NASA Keck PI program, the Carnegie
Institution of Washington, and the NAI, NASA Astrobiology Institute.
G.L. acknowledges support from NSF AST-0449986.  G.W.H. acknowledges support from
NASA, NSF, Tennessee State University, and the state of Tennessee through its
Centers of Excellence program. The work herein is based on observations obtained
at the W. M. Keck Observatory, which is operated jointly by the University of
California and the California Institute of Technology, and we thank the UC-Keck
and NASA-Keck Time Assignment Committees for their support.  This research has
also made use of the Keck Observatory Archive (KOA), which is operated by the
W. M. Keck Observatory and the NASA Exoplanet Science Institute (NExScI), under
contract with the National Aeronautics and Space Administration. We gratefully
acknowledge William Cochran for a 10-night dataset that has been obtained
through KOA and incorporated into this paper.  We also wish to extend our
special thanks to those of Hawaiian ancestry on whose sacred mountain of Mauna
Kea we are privileged to be guests. Without their generous hospitality, the Keck
observations presented herein would not have been possible. This research has
made use of the SIMBAD database, operated at CDS, Strasbourg, France.

{\it Facilities:} \facility{\telescope\ (\spectrograph)}.

\begin{deluxetable}{lrr}
\tabletypesize{\scriptsize}
\tablecolumns{3}
\tablewidth{0pt}
\tablecaption{Stellar Parameters for \starname}
\tablehead{
\colhead{Parameter} & \colhead{Value} & \colhead{Reference}\\
}
\startdata
\label{stellarparams}
Spec.~Type         & \spectype    & \citet{cen07}    \\
Mass (\msun)       & \starmassvaf & \citet{vaf05}    \\
                   & \starmasstak & \citet{takeda07} \\
                   & \starmasssou & \citet{sou08}    \\
Radius (\rsun)     & \starradvaf  & \citet{vaf05}    \\
                   & \starradtak  & \citet{takeda07} \\
Luminosity (\lsun) & \starlumvaf  & \citet{vaf05}    \\
                   & \starlumsou  & \citet{sou08}    \\
Distance (pc)      & \distance    & \citet{per97}    \\
$V\sin{i}$ (\kms)  & \vsini       & \citet{vaf05}    \\
$\log{R'_{\rm HK}}$ & \rhkjtw      & \citet{jtw04}    \\
                   & \rhkhalla    & \citet{hall07}   \\
                   & \rhkhallb    & \citet{hall09}   \\
                   & \rhkours     & This work        \\
$P_{\rm rot}$ (days) & \prot       & \citet{jtw04}    \\
age (Gyr)           & \agejtw     & \citet{jtw04}    \\
                    & \agevaf     & \citet{vaf05}    \\
                    & \agetak     & \citet{takeda07} \\
$[\rm Fe/H]$        & \fehvaf     & \citet{vaf05}    \\
                    & \fehcen     & \citet{cen07}    \\
                    & \fehsou     & \citet{sou08}    \\
$T_{\rm eff}$ (K)    & \teffsou    & \citet{vaf05,sou08} \\
                    & \teffcen    & \citet{cen07}    \\
$\log{g}$           & \loggtak    & \citet{vaf05,takeda07} \\
                    & \loggcen    & \citet{cen07}    \\
                    & \loggsou    & \citet{sou08}    \\
\enddata
\end{deluxetable}

\clearpage
\begin{deluxetable}{rrrr}
\tabletypesize{\scriptsize}
\tablecaption{Radial Velocities for \starname}
\tablewidth{0pt}
\tablehead{
Barycentric JD & RV & Error & Observatory \\
(-2450000)   &  (\ms) & (\ms) & 
}
\startdata
\label{vels}
 366.93641 &  -0.21 & 1.36 & K \\
 715.00792 &  -9.17 & 1.11 & K \\
 785.76804 & -14.14 & 1.22 & K \\
1010.09403 &  -5.98 & 1.17 & K \\
1068.92515 &  -5.81 & 1.14 & K \\
1173.72568 &  -0.48 & 1.10 & K \\
1343.08899 &  -2.04 & 1.12 & K \\
1374.10891 &  -4.86 & 1.27 & K \\
1438.88093 &  -3.75 & 1.55 & K \\
1550.73862 &  -4.91 & 1.25 & K \\
1883.80712 &  -0.32 & 1.34 & K \\
2098.12817 &  -7.96 & 1.51 & K \\
2236.72882 &  -0.08 & 1.30 & K \\
2489.05617 &  -2.67 & 1.51 & K \\
2537.92834 &  -3.48 & 1.57 & K \\
2575.84461 &  -1.98 & 1.54 & K \\
2829.05914 &  -7.47 & 1.43 & K \\
2899.03253 &  -2.51 & 1.37 & K \\
3196.06321 &   2.86 & 1.28 & K \\
3238.01691 &   0.50 & 0.21 & K \\
3238.90057 &   2.88 & 0.42 & K \\
3239.94328 &   6.80 & 0.44 & K \\
3240.99951 &   6.49 & 0.49 & K \\
3241.11480 &   7.06 & 0.65 & K \\
3301.85378 &  -4.94 & 0.65 & K \\
3302.81034 &  -0.65 & 0.61 & K \\
3338.75610 &   1.71 & 1.08 & K \\
3339.75947 &  -1.18 & 0.66 & K \\
3369.75405 &  -6.98 & 0.64 & K \\
3397.70379 &   3.24 & 0.58 & K \\
3398.73620 &  -0.30 & 0.61 & K \\
3399.69869 &  -2.37 & 0.64 & K \\
3400.70946 &   4.84 & 0.64 & K \\
3547.12033 &   3.03 & 0.63 & K \\
3548.10688 &   3.59 & 0.96 & K \\
3549.13414 &   0.06 & 1.14 & K \\
3550.11494 &   3.57 & 0.61 & K \\
3551.12904 &   4.00 & 0.62 & K \\
3552.12582 &   4.68 & 0.58 & K \\
3603.04546 &   7.42 & 0.63 & K \\
3604.06836 &  10.32 & 0.75 & K \\
3605.04697 &   6.44 & 0.64 & K \\
3692.92330 &  -2.41 & 0.61 & K \\
3693.84499 &   1.79 & 0.54 & K \\
3694.78313 &   1.37 & 0.58 & K \\
3695.70050 &   4.03 & 0.53 & K \\
3695.82845 &   4.85 & 0.59 & K \\
3696.76371 &   5.63 & 0.57 & K \\
3723.82034 &  -1.81 & 0.60 & K \\
3724.74409 &   1.77 & 0.63 & K \\
3746.73490 &  -1.19 & 0.69 & K \\
3747.79465 &  -0.46 & 0.97 & K \\
3748.72482 &   1.91 & 0.68 & K \\
3749.72744 &  -0.78 & 0.62 & K \\
3775.71584 &  -2.90 & 0.67 & K \\
3776.70867 &   2.10 & 0.65 & K \\
3777.71251 &   1.53 & 0.80 & K \\
3778.71106 &  -2.19 & 0.86 & K \\
3779.73772 &  -2.38 & 0.81 & K \\
3927.10455 &   0.42 & 0.62 & K \\
3928.05300 &   1.31 & 0.85 & K \\
3959.10907 &  -1.33 & 0.62 & K \\
3961.04337 &   0.02 & 0.64 & K \\
3962.01510 &   2.99 & 0.61 & K \\
3981.90046 &  -3.83 & 0.65 & K \\
3982.98884 &   2.04 & 0.66 & K \\
3983.87130 &   2.26 & 0.58 & K \\
3984.97007 &   6.59 & 0.65 & K \\
4083.81187 &   2.61 & 0.73 & K \\
4084.77778 &   3.39 & 0.73 & K \\
4129.74967 &   8.16 & 0.66 & K \\
4131.71246 &   4.76 & 0.75 & K \\
4279.11113 &   0.46 & 1.19 & K \\
4280.10709 &  -1.69 & 1.26 & K \\
4286.10589 &  -5.34 & 1.22 & K \\
4295.08454 &  -2.01 & 1.15 & K \\
4305.06828 &  -5.52 & 0.61 & K \\
4306.02305 &  -1.97 & 0.83 & K \\
4306.99662 &  -3.53 & 0.94 & K \\
4307.10265 &  -0.77 & 0.83 & K \\
4308.07916 &  -0.27 & 0.71 & K \\
4309.07893 &   0.58 & 0.85 & K \\
4310.02797 &  -4.99 & 0.80 & K \\
4310.12123 &  -2.37 & 1.19 & K \\
4311.00946 &  -6.55 & 0.84 & K \\
4311.10912 &  -2.30 & 1.23 & K \\
4312.00404 &  -4.29 & 0.82 & K \\
4312.11147 &  -2.84 & 0.82 & K \\
4313.00139 &  -3.43 & 0.87 & K \\
4313.10575 &  -2.31 & 0.86 & K \\
4313.99949 &  -0.07 & 0.85 & K \\
4314.10977 &   0.60 & 0.90 & K \\
4315.12175 &  -1.56 & 0.86 & K \\
4319.01064 &   3.70 & 0.70 & K \\
4336.05526 &   1.63 & 0.75 & K \\
4337.10146 &   4.33 & 0.81 & K \\
4343.93370 &   5.91 & 1.21 & K \\
4396.76696 &  -2.48 & 1.18 & K \\
4397.82001 &  -4.10 & 1.23 & K \\
4398.84808 &  -2.28 & 1.12 & K \\
4427.86323 &   0.36 & 0.67 & K \\
4429.77428 &   2.62 & 1.30 & K \\
4430.77076 &   2.35 & 1.24 & K \\
4454.82508 &  -2.21 & 1.20 & K \\
4456.80064 &  -2.24 & 1.25 & K \\
4460.78595 &  -4.89 & 1.35 & K \\
4461.79834 &  -6.06 & 1.13 & K \\
4464.80653 &   4.89 & 1.37 & K \\
4492.70875 &   5.97 & 1.29 & K \\
4634.11567 &  -6.90 & 0.92 & K \\
4635.07070 &  -5.12 & 0.88 & K \\
4636.09364 &  -3.98 & 0.87 & K \\
4637.10870 &  -6.08 & 0.88 & K \\
4638.11047 &  -2.74 & 0.91 & K \\
4639.09420 &  -4.49 & 0.88 & K \\
4640.10790 &  -4.80 & 0.84 & K \\
4641.10062 &  -7.37 & 0.86 & K \\
4642.07466 &   1.06 & 0.94 & K \\
4644.12501 &   1.89 & 0.89 & K \\
4717.94972 &  -2.96 & 0.83 & K \\
4719.00539 &   0.38 & 0.73 & K \\
4720.00787 &  -4.48 & 0.66 & K \\
4720.96693 &  -4.91 & 0.77 & K \\
4721.98581 &  -5.73 & 0.79 & K \\
4722.89844 &  -0.54 & 0.71 & K \\
4723.96804 &   0.62 & 0.73 & K \\
4724.96948 &  -0.12 & 0.82 & K \\
5022.12042 &  -5.09 & 0.52 & K \\
5023.08514 &  -3.59 & 0.85 & K \\
5024.11008 &  -1.22 & 0.49 & K \\
5025.12159 &  -4.13 & 0.58 & K \\
5050.06798 &  -2.46 & 0.49 & K \\
5051.09955 &  -2.43 & 0.46 & K \\
5052.04858 &   1.40 & 0.55 & K \\
5053.08104 &  -0.45 & 0.56 & K \\
5054.04263 &   4.05 & 0.59 & K \\
3713.70121 &   0.45 & 0.70 & Q01 \\
3713.76642 &   1.42 & 0.59 & Q01 \\
3713.82984 &   3.44 & 0.60 & Q01 \\
3714.69699 &   0.24 & 0.61 & Q01 \\
3714.75572 &   0.56 & 0.58 & Q01 \\
3714.82460 &   0.22 & 0.56 & Q01 \\
3715.68459 &  -1.65 & 0.66 & Q01 \\
3715.72791 &   0.10 & 0.70 & Q01 \\
3715.77573 &  -2.24 & 0.64 & Q01 \\
3715.82808 &  -0.68 & 0.63 & Q01 \\
3716.72884 &  -3.10 & 0.59 & Q01 \\
3716.77010 &  -1.49 & 0.63 & Q01 \\
3716.82182 &  -1.79 & 0.72 & Q01 \\
3717.79314 &  -4.50 & 0.56 & Q01 \\
3726.72138 &  -1.00 & 0.61 & Q01 \\
3726.77494 &   2.24 & 0.59 & Q01 \\
3727.71647 &  -1.39 & 0.54 & Q01 \\
3727.77202 &  -4.18 & 0.56 & Q01 \\
3727.81049 &  -2.38 & 0.58 & Q01 \\
3728.73194 &  -5.62 & 0.54 & Q01 \\
3728.79946 &  -2.87 & 0.59 & Q01 \\
3728.84615 &  -2.39 & 0.58 & Q01 \\
3728.84971 &   0.18 & 0.60 & Q01 \\
3729.70727 &   1.63 & 0.85 & Q01 \\
3729.81947 &   0.00 & 0.60 & Q01 \\
3730.71913 &   2.23 & 0.60 & Q01 \\
3730.77245 &   2.82 & 0.56 & Q01 \\
3730.80874 &   2.49 & 0.54 & Q01 \\
\enddata
\end{deluxetable}

\begin{deluxetable}{lr@{$\pm$}lr@{$\pm$}lccr@{$\pm$}lr@{$\pm$}lr@{$\pm$}l}
\tabletypesize{\scriptsize}
\tablecolumns{13}
\tablewidth{0pt}
\tablecaption{Circular Solutions (Epoch \fitepoch) \label{planetparamsnoe}}
\tablehead{
\colhead{Planet} & \multicolumn{2}{c}{Period} & \multicolumn{2}{c}{$K$}     & \colhead{$e$} & \colhead{$\omega$}  & \multicolumn{2}{c}{$M$}       & \multicolumn{2}{c}{$M\sin{i}$}    & \multicolumn{2}{c}{$a$} \\
\colhead{}       & \multicolumn{2}{c}{(d)} & \multicolumn{2}{c}{(\ms)} & {}            & {(deg)}         & \multicolumn{2}{c}{(deg)} & \multicolumn{2}{c}{($M_{\oplus}$)} & \multicolumn{2}{c}{(AU)}
}
\startdata
\cutinhead{1-planet; $\chi_{\nu}^2=19.249$; rms=3.43\,\ms}
HD1461 b & 5.7726 & 0.0026 & 2.6 & 0.4 & 0.0 & n/a &  79 & 60 & 7.4 & 1.2 & 0.063437 & 0.000019\\
\cutinhead{2-planets; $\chi_{\nu}^2=13.151$; rms=2.87\,\ms}
HD1461 b & 5.7718 & 0.0010 & 2.8 & 0.3 & 0.0 & n/a &  44 & 36 &  7.8 &  0.8 & 0.063431 & 0.000008\\
HD1461 d & 4019   & 1433   & 2.8 & 1.9 & 0.0 & n/a & 285 & 47 & 69.9 & 78.9 & 4.98     & 0.99    \\
\cutinhead{3-planets; $\chi_{\nu}^2=10.093$; rms=2.41\,\ms}
HD1461 b & 5.7718 & 0.0010 & 2.7 & 0.2 & 0.0 & n/a &  45 & 37 &  7.6 &  0.7 & 0.063431 & 0.000008\\
HD1461 c & 446    & 9      & 2.3 & 0.4 & 0.0 & n/a &  82 & 52 & 27.9 &  4.9 & 1.151    & 0.016   \\
HD1461 d & 5017   & 1171   & 3.2 & 1.4 & 0.0 & n/a & 319 & 36 & 87.1 & 51.8 & 5.78     & 0.82    \\
\enddata
\end{deluxetable}

\begin{deluxetable}{lr@{$\pm$}lr@{$\pm$}lr@{$\pm$}lr@{$\pm$}lr@{$\pm$}lr@{$\pm$}lr@{$\pm$}l}
\tabletypesize{\scriptsize}
\tablecolumns{15}
\tablewidth{0pt}
\tablecaption{Eccentric Solutions (Epoch \fitepoch) \label{planetparamswithe}}
\tablehead{
\colhead{Planet} & \multicolumn{2}{c}{Period} & \multicolumn{2}{c}{$K$}     & \multicolumn{2}{c}{$e$} & \multicolumn{2}{c}{$\omega$}  & \multicolumn{2}{c}{$M$}       & \multicolumn{2}{c}{$M\sin{i}$}    & \multicolumn{2}{c}{$a$} \\
\colhead{}       & \multicolumn{2}{c}{(d)} & \multicolumn{2}{c}{(\ms)} & \multicolumn{2}{c}{}    & \multicolumn{2}{c}{(deg)} & \multicolumn{2}{c}{(deg)} & \multicolumn{2}{c}{($M_{\oplus}$)} & \multicolumn{2}{c}{(AU)}
}
\startdata
\cutinhead{1-planet; $\chi_{\nu}^2=19.349$; rms=3.43\,\ms}
HD1461 b & 5.7727 & 0.0025 & 2.7 & 0.6 & 0.14 & 0.19 &  58 & 69 &  26 & 71 & 7.6 & 1.2 & 0.063438 & 0.000018\\
\cutinhead{2-planets; $\chi_{\nu}^2=12.768$; rms=2.64\,\ms}
HD1461 b & 5.7720 & 0.0011 & 2.8 & 0.4 & 0.06 & 0.14 &  28 & 70 &  24 & 67 &  8.1 &  0.9 & 0.063432 & 0.000008\\
HD1461 d & 7000   & 200000 & 4   & 24  & 0.51 & 0.24 & 319 & 32 & 210 & 83 &  101 &  357 & 7        & 42      \\
\cutinhead{3-planets; $\chi_{\nu}^2=8.983$; rms=2.24\,\ms}
HD1461 b & 5.7722 & 0.0011 & 2.8 & 0.3 & 0.04 & 0.01 & 186 & 63 & 234 & 77 &  8.1 &  0.7 & 0.063434 & 0.000008\\
HD1461 c & 454    & 4      & 2.8 & 2.7 & 0.74 & 0.13 &  87 & 28 & 237 & 28 & 22.9 &  9.8 & 1.165    & 0.008   \\
HD1461 d & 5000   & 90000  & 4   &  70 & 0.16 & 0.29 & 326 & 79 & 136 & 85 &   97 & 1161 & 5        & 18      \\
\enddata
\end{deluxetable}

\end{document}